\documentstyle[aps,multicol,epsfig,psfig]{revtex}   

\newcommand{\beq}{\begin{equation}}  
\newcommand{\eeq}{\end{equation}}  
\newcommand{\bea}{\begin{eqnarray}}  
\newcommand{\eea}{\end{eqnarray}}  
  
\begin{document}     
\title{Classical double ionization of atoms in strong fields}
\author{
Bruno Eckhardt$^1$ and Krzysztof Sacha$^{1,2}$
}     
\address{$^1$ Fachbereich Physik, Philipps Universit\"at    
	 Marburg, D-35032 Marburg, Germany}    
\address{$^2$ Instytut Fizyki im. Mariana Smoluchowskiego,
Uniwersytet Jagiello\'nski, ul. Reymonta 4,
PL-30-059 Krak\'ow, Poland}
\date{\today}     
\maketitle{ }   
  
\begin{abstract}
Recent high resolution measurements of the momenta of 
two electrons emitted from Argon atoms in a strong laser field show
a strong preference for the outgoing electrons to have similar
momenta and to be ejected in the same direction along the field. 
We discuss the final stages of this process in a classical model
in the spirit of Wanniers approach to double ionization after
electron impact.
Starting from an intermediate state of high but negative energy the 
field opens up a channel through which both electrons can escape. Near 
the threshold for this process Coulomb repulsion favors an 
escape with the electron momenta and positions
symmetric with respect to the electric field axis. 
Classical trajectory simulations
within this symmetry subspace account for most features of the 
observed momentum distribution. 
\end{abstract}     
  
\begin{multicols}{2}   
Double ionization of neutral atoms (in particular
He, Ne and Ar) in the presence of strong fields has attracted
considerable attention after it was noted that the observed
yield is much higher than can be expected on the basis
of an independent electron model \cite{Kulander1,Kulander2}. It was
proposed early on that correlations between the electrons
should be responsible for this enhancement. A striking demonstration
of such correlations is provided by recent high resolution 
experiments on the distribution of the ion-recoil momenta 
\cite{weber1,weber2,rottke} and the electron momenta
\cite{weber3} which clearly show
a preference for the symmetric escape of both electrons to the same side
of the nucleus. 
This is very different from the process in the absence
of a field (Wannier-ionization, \cite{Wannier,Rau})  
where the dominant ionization path
has both electrons escape symmetrically placed on opposite sides 
of the nucleus. In particular, while the Wannier mode minimizes the
electron repulsion, a symmetric escape to the same side has considerable
residual energy in the repulsion between electrons.
So why is this channel favoured?

It is by now generally accepted that the ionization takes 
place in two steps: a high excitation of one electron
followed by the double ionization event
\cite{Corkum,Kulander3,Kulander4,becker1,becker2,Kulander5,becker3,becker4}. 
The highly excited electron
is accelerated by the field and driven back to the core where it
collides with the other electron, and transfers enough energy
and momentum so that both electrons can escape from the nucleus. 
It is this rescattering of the electron that enhances the energy
transfer and thus the ionization rate. But it also implies that 
during the ionization process the interaction between the
electrons cannot be ignored.

The amount of energy the excited electron can gain during half a 
period of the electric field has been estimated in 
\cite{weber1}: for a field intensity of $2.9\cdot 10^{14} W/cm^2$
the estimate gives a maximal energy transfer of about $60 eV$.
This is insufficient to ionize the second electron directly, but
as the experiments show, double ionization is possible nevertheless. 
This implies that there must be a mechanism by which the electrons can 
draw additional energy from the field so that asymptotically, once the 
pulse has ceased, the total energy of the system is positive.

The related problem of double ionization in collisions or
single photon excitation was studied in a classic paper
by Wannier \cite{Wannier,Rau}.
He assumed that during the first stages a high
energy complex of electrons close to the nucleus is formed
from which then the ionized electrons escape. If the energy is close
to threshold, they cannot afford to put energy into the mutual
repulsion and the escape is with both electrons on opposite
sides of the nucleus. Moreover, he argued that their distance has
to be the same, for any difference in position and energy would be
amplified, pushing the configuration towards single electron
ionization.

Essential elements of this discussion also apply to the present
situation. At the end of the first step the rescattering of one electron 
to the core produces a highly excited two electron complex. The estimates
show \cite{weber1} that the total energy in the system is 
insuffient to doubleionize
immediately. However, if the electric field during the collision is
non-zero, a saddle opens through which the electrons can escape.
Because of their mutual repulsion the electron that gets to the 
saddle first has an advantage: 
it can cross the saddle while pushing the other back to the nucleus.
Since energy is scarce near threshold this results in either
single ionization or in another rescattering event, but most likely
not in double ionization. This suggests that the dominant path leading
to double ionization has both electrons cross the saddle
side by side. This singular process then acquires a finite probability
when trajectories asymptotic to this configuration are taken into
account.

Therefore, we propose that near the threshold for double ionization
the only path leading to double ionization has both 
electrons escape symmetric with respect to the field axis.
With the field pointing along the $x$-axis and
the electrons confined to the plane $z=0$ their coordinates are
$(x,y,0)$ and $(x,-y,0)$ in position and $(p_x, p_y, 0)$ and
$(p_x, -p_y, 0)$ in momenta. In this geometry the observation 
that both electrons escape to the same side is built from the outset.
As an aside we note that this symmetry plane also 
contains the Wannier orbit, for $x=0$, as the symmetric escape 
perpendicular to the field axis.

The classical Hamilton function for this geometry then is
(in atomic units, with infinitely heavy nucleus and in 
dipole approximation)
\beq
H(p_x, p_y, x, y, t)= p_x^2+p_y^2 + V(x,y,t)
\eeq
with potential energy
\beq
V(x,y,t) = 
- \frac{4}{\sqrt{x^2+y^2}}
+ \frac{1}{2y} + 2 F\, x\, f(t)\, \cos(\omega t+\phi)
\eeq
and the pulse shape
\beq
f(t)=\sin^2(\pi t/T_d) 
\eeq
where the duration of the pulse is taken to be four field cycles,
$T_d=8\pi /\omega$. The frequency is $\omega/2=0.057\,a.u.$,
and corresponds to the experimental situation \cite{weber1,weber2,weber3}.
The rescattering of the electrons leads to a highly excited complex
of total energy $\tilde E$ which every now and then is close to the 
symmetric configuration described by the Hamiltonian (1). Any configuration
on this energy shell (for some fixed time $t$) as well as any phase 
$\phi$ of the field is equally likely, and the experimental 
observations are averages over initial conditions and phases.

As mentioned, for the weakest fields where double ionization is
observed the rescattered electron does not bring in enough
energy for double ionization. However, if the collision happens
near a time where the field is strong, the electric field distorts
the potential and opens a path for escape in down-field direction.
This process can be discussed adiabatically for fixed external
field since the motion of the electrons near the nucleus is 
much faster than the change in the field. The ionization can 
thus be discussed in the potential (2) with fixed field.
Equipotential lines for the potential (2) at a maximum
of the field for $F=0.137\, a.u.$, corresponding to 
an intensity of $6.6\cdot 10^{14} W/cm^2$, are shown in Fig.~1. 
The saddle is located
along the line $x=r_S \cos \theta$ and $y=r_S \sin \theta$ with
$\theta = \pi/6$ or $5\pi/6$ and at a distance
$r_S^2=\sqrt{3}/F_{max}$ where 
$F_{max}=\mbox{max\,}_t |F\, f(t)\, \cos(\omega t +\phi)|$. 
The energy of the saddle is 
\beq
V_S = - 6 \sqrt{F_{max}/\sqrt{3}}\,.
\eeq
For the extremal fields in a pulse of the above mentioned intensity 
this gives $V_S=-1.69\, a.u.$, so that within the adiabatic picture
the saddle can be reached if the returning electron brings
in at least $1.22\, a.u.$ in energy. During a field cycle the saddle
moves in from infinity along the line at $\theta=\pi/6$, moves out again
to infinity after half a period and then moves in and out again
along the line $\theta=5\pi/6$ during the second half of the cycle.
Ionization is most likely when the saddle is closest to te nucleus.
%

A typical trajectory within the symmetric configuration is shown in
Fig.~\ref{traj}. During the ramping of the field the electronic motion
is little influenced by the electric field, but during the third half cycle
of the field the saddle is close enough to the electron orbits and 
ionization takes place. Once on the other side of the saddle,
the electrons rapidly gain energy.
The saddle thus provides a kind of transition state \cite{Wigner,Pollak}
for the double ionization process: once the electrons cross it, 
they are accelerated by the field and pulled further away, making
a return rather unlikely. Moreover, they can aquire the missing 
energy so that both electrons can escape even when the field vanishes. 
The field thus plays a double role in determining a threshold
for this process: during the first stages of the rescattering process 
it provides the energy for the collision complex and during the 
final stages it opens the path for double escape.

In the experiments it is not possible to monitor all details and
intermediate stages of the process. Most information is extracted
from the distribution of final momenta ${\bf p}_n$ of the nucleus
and ${\bf p_i}$ of the electrons, where
${\bf p}_1 + {\bf p}_2 \approx - {\bf p}_n$ 
\cite{weber1,weber2,weber3,rottke}. 
Because of the symmetry assumption in the model the components
perpendicular to the field vanish. The distribution of the parallel
components can be calculated by averaging over all initial conditions of 
prescribed energy and all phases of the field.

Classical scaling of the Hamiltonian (1) implies that the results
do not depend on the initial energy $\tilde E$ and field strength 
independently, but on the combination $F/{\tilde E}^2$ only. The field
strength $F$ is set by the intensity of the laser. The initial energy 
of the two-electron complex is determined by the
field dependent efficieny of the single electron excitation step
and thus not directly accessible, although it can be estimated
as in \cite{weber1}. We therefore fix $F$ and vary
initial energy.
The results for fixed field strength $F=0.137\, a.u.$ and an initial
energy of $\tilde E=-0.58\, a.u.$ are
compared to the experimental distribution in Fig.~\ref{distrib}. 
The final distribution of momenta clearly shows the double hump
structure indicating a preference for ionization
parallel and antiparallel to the field. The maxima of
the experimental distribution are at about $p_\parallel=\pm 1.5\, a.u.$, 
whereas
the numerical ones within the symmetric subspace lie at 
 about $\pm 1\, a.u.$. We take this close agreement as strong indication
that double ionization can only occur in the neighborhood of the symmetric
process discussed here.

For lower energy, $\tilde E=-1.3\, a.u.$ the minimum
at $p_\parallel=0$ almost vanishes and only a single maximum shows up 
(Fig.~\ref{distrib2}a).
This corresponds to the experimental situation of a weaker pulse
which evidently transfers less energy to the rescattered electron
\cite{weber1}. Within the symmetric subspace,
there are two reasons for this change: at this lower initial 
energy the electrons have little kinetic energy when crossing the
barrier, so that the splitting should be expected to be small.
Secondly, the electrons cross the barrier typically when the field
is strongest, so that after the ionization there 
is still considerable smearing of the distribution due to 
the interaction with the remainder of the pulse. 
If the distribution is monitored
immediately after the crossing of the barrier the two preferred
momenta parallel and anti-parallel to the field stand out 
clearly, as demonstrated in Fig.~\ref{distrib2}b. 

These numerical results in the reduced symmetry subspace are in 
surprising agreement with the experimental data. They show that
the configuration with rotational symmetry around the field
axis dominates the cross section. The interaction with the
field shortly after ionization is responsible for most of the
smearing of the final distribution, additional contributions
come from trajectories that are not symmetric. Actually,
non-symmetric configurations are needed in order to obtain
a non-vanishing cross section to begin with: the symmetric subspace
is a set of measure zero in the phase space of the six
degree of freedom system and acquires a finite overlap with initial
conditions only due to non-symmetric initial conditions that 
are asymptotic to the symmetric subspace. The 
cross section for double ionization shows that this overlap is 
small \cite{weber1,weber2,weber3,rottke}.

The picture proposed here for multiphoton double ionization is very
similar to that of Wannier for double ionization through electron
impact. The main difference is that now the symmetry between the
outgoing electrons is a rotation around the field axis whereas it
is a point symmetry with the nucleus in the center in Wanniers case.

Finally, we would like to mention that besides the different symmetry
there is another difference to 
the Wannier double ionization without field: In the zero field 
ionization there is only a single
trajectory so that there in leading order semiclassical approximation
no quantum interferences can be expected\cite{Rost}. 
The present problem falls into the category of quantum chaotic 
scattering \cite{Eckhardt,Smilansky} where the classical
ionization dynamics is chaotic and quantum interference effects
between different paths cannot be ruled out. It will be interesting
to pinpoint quantum interference effects in this system.

We would like to thank Harald Giessen for stimulating our interest
in this problem and for discussions of the experiments.
Financial support by the Alexander von Humboldt
Foundation and by KBN under project 2P302B00915
are gratefully acknowledged.



\newpage

\narrowtext

\begin{figure}
\epsfig{file=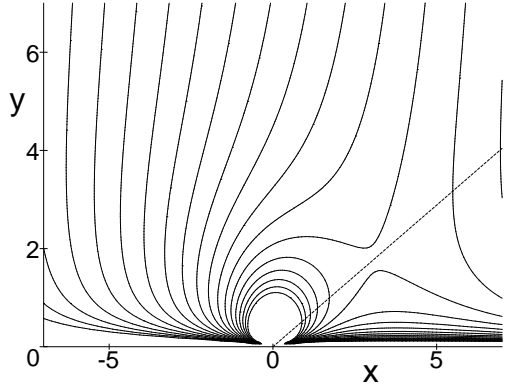%
,scale=0.3,angle=-90
}
\caption[]{
Adiabatic potential $V(x,y,t)$ for fixed time $t$ 
in the symmetric subspace. The saddle moves along the 
dashed line when the electric field points in the positive
$x$-direction and along a second obtained by reflection on
$x=0$ during the other half of the field cycle. 
}\label{phspace}
\end{figure}


\begin{figure}
\epsfig{file=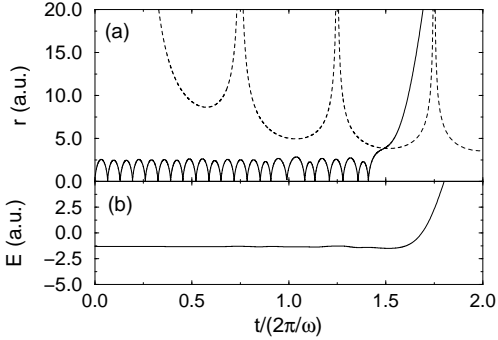%
,scale=0.3,angle=-90
}
\caption[]{
A typical trajectory in the symmetric subspace with $\tilde E=-1.3\, a.u.$ 
(a) distance of the electrons to
the nucleus. The dashed line indicates the distance
of the saddle. Note that before the double ionization occurs the
effect of the field on the electrons is minimal, supporting the adiabatic
assumption.
(b) energy of the electrons. 
Note that the initial state has negative total energy and cannot
lead to double ionization. The energy increases once the electrons
have escaped from the nucleus far enough so that acceleration by
the electric field dominates.
}\label{traj}
\end{figure}

\vskip10cm

\begin{figure}
\psfig{file=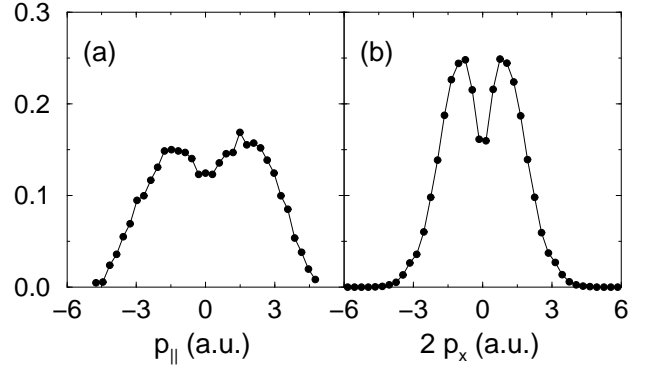%
,scale=0.4,angle=-90
}
\caption[]{
Final distribution of ion momenta parallel to the field for
$F=0.136\, a.u.$: (a) experimental distribution from \cite{weber1}
(b) distribution from symmetric subspace with initial energy 
$\tilde E=-0.58\, a.u.$. The classical distribution is based
on an ensemble of $2\cdot10^5$ trajectories.
}\label{distrib}
\end{figure}

\begin{figure}
\psfig{file=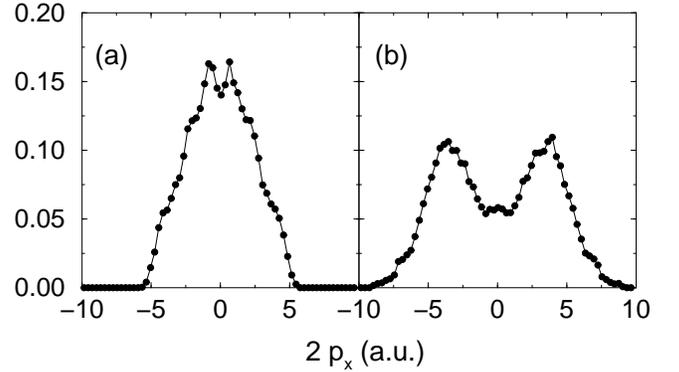%
,scale=0.4,angle=-90
}
\caption[]{
Final distribution of momenta parallel to the field for
$F=0.136\, a.u.$ and initial energy $\tilde E=-1.3\, a.u.$. 
(a) at the end of the pulse; 
(b) at the first zero of the field after crossing the saddle.
This figure demonstrates the smearing of the distribution due to the
final post ionization interaction with the electric field.
It is based on 50.000 classical trajectories.
}\label{distrib2}
\end{figure}

\vfill
\end{multicols}

\begin{references}  

\bibitem{Kulander1} D.N. Fittinghof, P.R. Bolton, B. Chang,
and K.C. Kulander, Phys. Rev. Lett. {\bf 69}, 2642 (1992)

\bibitem{Kulander2} B. Walker, B. Sheehy, L.F. DiMauro, P. Agostini,
K.J. Schafer, and K.C. Kulander, Phys. Rev. Lett. {\bf 73}, 1227 (1994)

\bibitem{weber1} Th. Weber, M. Weckenbrock, A. Staudte, L. Spielberger, 
O. Jagutzki, V. Mergel, F. Afaneh, G. Urbasch, M. Vollmer, H. 
Giessen and R. D\"orner, Phys. Rev. Lett. {\bf 84}, 443 (2000)

\bibitem{weber2} Th. Weber, M. Weckenbrock, A. Staudte, L. Spielberger, 
O. Jagutzki, V. Mergel, F. Afaneh, G. Urbasch, M. Vollmer, H. 
Giessen and R. D\"orner, J. Phys. B: At. Mol. Opt. Phys. 
{\bf 33}, L1 (2000)

\bibitem{weber3} Th. Weber, H. Giessen, M. Weckenbrock, G. Urbasch,
A. Staudte, L. Spielberger, 
O. Jagutzki, V. Mergel, M. Vollmer, and R. D\"orner, 
Nature {\bf ???}, ??? (2000) (in press)

\bibitem{rottke}
R. Moshammer, B. Feuerstein, W. SChmitt, A. Dorn, C.D. Sch\"oter,
J. Ullrich, H. Rottke, C. Trump, M. Wittmann, G. Korn, K. Hoffmann
and W. Sandner, Phys. Rev. Lett. {\bf 84}, 447 (2000)

\bibitem{Wannier}  
{G.H. Wannier}, {Phys. Rev.} {\bf 90}, {817} (1953)

\bibitem{Rau}  
{A. R. P. Rau }, {Phys. Rep.} {\bf 110}, {369}, (1984)

\bibitem{Corkum} P.B. Corkum, Phys. Rev. Lett. {\bf 71}, 1994 (1993)

\bibitem{Kulander3} K.C. Kulander, J. Cooper, and K.J. Schafer,
	Phys. Rev. A {\bf 51}, 561 (1995)

\bibitem{Kulander4} B. Walker, B. Sheehy, K.C. Kulander, 
and L.F. DiMauro, Phys. Rev. Lett. {\bf 77}, 5031 (1996)

\bibitem{becker1} A. Becker and F.H.M. Faisal, 
J. Phys. B {\bf 29}, L197 (1996)

\bibitem{becker2} A. Becker and F.H.M. Faisal, 
J. Phys. B {\bf 32}, L335 (1999)

\bibitem{Kulander5} B. Sheehy, R. Lafon, M. Widmer, B. Walker,
L.F. DiMauro, P.A. Agostini,
and K.C. Kulander, Phys. Rev. A {\bf 58}, 3942 (1998)

\bibitem{becker3} A. Becker and F.H.M. Faisal, 
Phys. Rev. A {\bf 59}, R1742 (1999)

\bibitem{becker4} A. Becker and F.H.M. Faisal, 
Phys. Rev. Lett. {\bf ??}, ??? (2000) (in press)

\bibitem{Wigner} E. P. Wigner,
    Z. Phys. Chemie B \bf 19\rm, 203 (1932);
    Trans. Faraday Soc. \bf 34\rm 29, (1938)

\bibitem{Pollak} E. Pollak, in {\em Theory of Chemical Reactions}, 
	{vol III, M. Baer, ed., (CRC Press, Boca Raton, 1985}, p. 123

\bibitem{Rost} J.M. Rost, Phys. Rev. Lett. {\bf 72}, 1998 (1994);
Phys. Rep. {\bf 297}, 271 (1999)

\bibitem{Eckhardt} B. Eckhardt, Physica \bf D 33\rm, 89 (1988)

\bibitem{Smilansky} U. Smilansky.   Semiclassical Quantization of Chaotic 
Billiards -  A Scattering Approach. in  {\em Proc. of the  Les Houches 
Summer School on Mesoscopic Quantum Physics}. Elsevier Science Publ. (1995) 
Ed. E. Akkermans, G. Montambaux and J. L. Pichard. 

\end{references}
\end{document}